\documentclass[letterpaper,english,reprint,nofootinbib,aps,superscriptaddress,showpacs,showkeys]{revtex4-1}
\usepackage{babel,calc,amsmath,amsthm,amssymb,graphicx,subfigure,xcolor}
\usepackage[T1]{fontenc}
\usepackage[unicode=true]{hyperref}
\hypersetup{
     colorlinks=true,       		% false: boxed links; true: colored links
     linkcolor=blue,          	% color of internal links
     citecolor=red,            % color of links to bibliography
     urlcolor=magenta,           	% color of external links
 }
	
\begin{document}
	
%	\title{Raising and Lowering Operators for Jaynes-Cummings model}
\title{Solving the Jaynes-Cummings Model with Shift Operators Constructed by Means of the Matrix-Diagonalizing Technique}

	\author{Jie Zhou}
\affiliation{Theoretical Physics Division, Chern Institute of Mathematics, Nankai University, Tianjin 300071, People's Republic of China}

\author{Hong-Yi Su}
\email{hysu@mail.nankai.edu.cn}
\affiliation{Theoretical Physics Division, Chern Institute of Mathematics, Nankai University, Tianjin 300071, People's Republic of China}

	\author{ Fu-Lin Zhang}
	\email{flzhang@tju.edu.cn } \affiliation{Physics Department, School of Science, Tianjin University, Tianjin 300072, People's Republic of China}	

	\author{Hong-Biao Zhang}
\email{zhanghb017@nenu.edu.cn }\affiliation{Institute of Theoretical Physics,
		Northeast Normal University, Changchun 130024, People's Republic of
		China}

\author{Jing-Ling Chen}
\email{chenjl@nankai.edu.cn}
\affiliation{Theoretical Physics Division, Chern Institute of Mathematics, Nankai University, Tianjin 300071, People's Republic of China}

%	\author{Jie Zhou}
%\affiliation{Theoretical Physics Division, Chern Institute of Mathematics, Nankai University, Tianjin 300071, People's Republic of China}
%
%
%\author{Hong-Yi Su}
%\email{hysu@mail.nankai.edu.cn}
%\affiliation{Theoretical Physics Division, Chern Institute of Mathematics, Nankai University, Tianjin 300071, People's Republic of China}
%
%	\author{ Fu-lin Zhang}
%	\email{flzhang@tju.edu.cn } \affiliation{Physics Department, School of Science, Tianjin University, Tianjin 300072, People's Republic of China}	
%
%\author{Jing-Ling Chen}
%\email{chenjl@nankai.edu.cn}
%\affiliation{Theoretical Physics Division, Chern Institute of Mathematics, Nankai University, Tianjin 300071, People's Republic of China}
%
%	\author{Hong-Biao Zhang}
%	\email{zhanghb017@nenu.edu.cn } \affiliation{Institute of Theoretical Physics,
%		Northeast Normal University, Changchun 130024, People's Republic of
%		China}
%

%	\begin{abstract}
%		The Jaynes-Cummings model is solved with the raising and lowering (shift) operators, to get which the matrix-diagonalizing technique is used, and with which a quantum dynamical algebra is constructed. Thus a quantum dynamical symmetry is revealed to exist in the model. Bell nonlocality is also found present universally in the excitations states.
%	\end{abstract}

	\begin{abstract}
		The Jaynes-Cummings model is solved with the raising and lowering (shift) operators by using the
matrix-diagonalizing technique.  Bell nonlocality is also found present ubiquitously in the excitations states of the model.
	\end{abstract}

	
 \maketitle
	
\section {Introduction}	
Many quantum mechanical models with various interactions and potentials were conventionally addressed by solving their wave equations in a certain position or momentum coordinate, with account of variables separation, boundary conditions, single-valuedness, \emph{etc}. Such a way of solving, while highly worthwhile in obtaining explicit energy spectra and wavefunctions, can sometimes obscure the underlying symmetries of the quantum system that is considered. In comparison, operator methods~\cite{shift} --- particularly ones involving Lie algebras~\cite{lie} --- not only simplify the problem solving in practice, but also provide more insight into the solutions of other related models that share similar symmetries, whatever stationary or dynamical, in most cases.

Out of many an algebraic method the one proposed in~\cite{ge2000} stands out by dealing with nonlinear deformation algebra~\cite{dq93}, which is generated from shift operators obtained by solving
\begin{equation}
  [H,\mathcal{X}]=\mathcal{X}\mathcal{G},\label{method}
\end{equation}
where $H$ denotes the Hamiltonian, $\mathcal{X}$ is a \emph{closed operator set}, and $\mathcal{G}$ is a matrix to be diagonalized (c.f. Eq.~(\ref{matrix-tech}) below, which is equivalent to (\ref{method}) by subtracting a term $H$ from $\mathcal{G}$). The linear $su(2)$ or $su(1,1)$ Lie algebra can then be written out with these shift operators, revealing a dynamical symmetry of the system~\cite{tolinear}.

Using this method in the present paper, we will solve the Jaynes-Cummings model (JCM) proposed by Jaynes and Cummings in 1963 \cite{JC}. The model describes the system of a two-level atom interacting with a quantized mode of an optical cavity, with or without the presence of light. Its applications range from atomic physics, quantum optics \cite{V.Vedral}, and solid-state quantum information circuits \cite{Irish}, both experimentally and theoretically.
The Hamiltonian reads	
		\begin{eqnarray}\label{1}
			H=\omega( a^\dag a+\frac{1}{2})+g(\sigma^+ a+\sigma^- a^\dag)+\Delta \sigma_z,
		\end{eqnarray}
where $\sigma_{x,y,z}$ are Pauli matrices, $2\Delta$ is the level splitting of the two level system, $a(a^\dagger)$ are the destruction (creation) operators of a single bosonic mode with freequency $\omega$, and $g$ is the coupled coefficient, and $\sigma^\pm=\frac{1}{2}(\sigma_x \pm i\sigma_y)$.
Here, the conservation of a quantity $C=a^\dagger a+\frac{1}{2}\sigma_z$, commuting with $H$, means that the state space can be broken down into infinite two-dimensional subspaces, each eigenstate can be labeled by $C=0,1,2\cdots$. There are two eigenstates in the two-dimensional subspace can be labeled by $+$ and $-$ \cite{Braak}.	

The paper is organized as  follows. In Sec. II, we construct the
raising and lowering operators for the Hamiltonian (\ref{1}). In Sec. III, we compute the energy spectrum and the wave functions of the physical system. In Sec. IV, we construct algebraic structure based on the raising and lowering operators. In Sec. V, we prpose a test of Bell's inequality with the excitation states of the JCM. Discussion is made in the last section.

\section {Raising and lowering operators of the Hamiltonian}
	
For any Hamiltonian operator $H$, if there are operators ${\hat {\cal
		L}}^{\pm}$ satisfying the following commutation relation
\begin{equation}
	\label{sh} [H,{\hat {\cal L}}^{\pm}]={\hat {\cal
			L}}^{\pm}f^{\pm}(H),
\end{equation}
then ${\hat {\cal L}}^+$ and ${\hat {\cal L}}^-$ are called the
raising and lowering operators of operator $H$, respectively \cite{HuChen,Chen,Chen1,WL,wl} . For
example, let ${\hat {\cal L}}^-=a$, ${\hat {\cal L}}^+=a^\dag$, and
$f^{\pm}(H)=\pm \hbar \omega$, Eq. (\ref{sh}) reduces to the usual
case of the quantum linear harmonic oscillator, for which $[H, a]=-a
\hbar\omega$, $[H, a^\dag]=a^\dag \hbar\omega$. We refer the readers who are
interested in the general definition of raising and lowering
operators to Refs. \cite{shift} and \cite{Chen}. Note that the explicit form of the raising and
lowering operators ${\hat {\cal L}}^{\pm}$ for a specific
Hamiltonian system need not be mutually adjoint \cite{shift}.

With the canonical commutation relations
 \begin{equation}\label{3}
 \begin{split}
   		&\left[a,a^\dagger\right]=1,\\
 		&\left[\sigma^+,\sigma^-\right]=\sigma_z,~~\left[\sigma^-,\sigma^+\right]=-\sigma_z,\\
 		&\left[\sigma^+,\sigma_z\right]=-2\sigma^+,~~\left[\sigma^-,\sigma_z\right]=2\sigma^-,
 \end{split}
 \end{equation}	
we have
 \begin{align}
 	&[H,\sigma_z]=-2g\sigma^+ a+2g\sigma^- a^\dagger,\label{4}\\
 	&[H,a^\dagger a]=g\sigma^+ a-g\sigma^- a^\dagger,\label{5}\\
	&[H,\sigma^+ a] = -\frac{g}{\omega}\sigma_z H+(\frac{g^2}{\omega}-\delta)\sigma^+ a-\frac{g^2}{\omega}\sigma^- a^\dagger-\frac{g\delta}{2\omega},\label{6}\\
 	&[H,\sigma^- a^\dagger] = \frac{g}{\omega}\sigma_z H+(\delta+\frac{g^2}{\omega})\sigma^- a^\dagger-\frac{g^2}{\omega}\sigma^+a+\frac{g\delta}{2\omega},\label{7}
 \end{align}	
with  $\delta=\omega-2\Delta$. From Eqs. (\ref{4}),  (\ref{5}),we can know that:
\begin{eqnarray}
\left[ {H,{a^\dag }a + \frac{1}{2}{\sigma _z}} \right] = 0
\end{eqnarray}
So,the operator \[C = {a^\dag }a + \frac{1}{2}{\sigma _z}\] is a conserved quantity.

From (\ref{5}), (\ref{6}) and (\ref{7}) can then be rewritten in the following form	
\begin{align}
& HX=XG,\label{matrix-tech}\\
&X=(\sigma^+ a,\sigma^- a^\dagger,\sigma_z,1)\nonumber,
\end{align}
where $X$ is the closed operator set~\cite{ge2000}, and $G$ is a $4\times4$ matrix
\begin{eqnarray}
G=\left(\begin{array}{cccc}H-\delta+\frac{g^2}{\omega}&-\frac{g^2}{\omega}&-2g&0\\-\frac{g^2}{\omega}&H+\delta+\frac{g^2}{\omega}&2g&0\\-\frac{g}{\omega}H&\frac{g}{\omega}H&H&0\\-\frac{g\delta}{2\omega}&\frac{g\delta}{2\omega}&0&H
\end{array}\right).
\end{eqnarray}
Solving the equation
    	\begin{eqnarray}\label{12}
    	\det(G-\lambda I)=0,
    	\end{eqnarray}
    	we obtain four eigenvalues of matrix $G$:    	
    	\begin{align}\label{13}
    	\lambda_1&=\lambda_2=H,\nonumber\\
    	\lambda_3&=\frac{g^2+H\omega-T(H)}{\omega},\nonumber\\
    	\lambda_4&=\frac{g^2+H\omega+T(H)}{\omega}.
    	\end{align}	
where $T(H)=\sqrt{g^4+4g^2 H\omega+\omega^2 \delta^2}$.	
	
	Now we can write $G$ as
	\begin{eqnarray}\label{14}
	G=R\Lambda R^{-1},
	\end{eqnarray}
and
	\begin{align}
	\Lambda&=\left(\begin{array}{cccc}\lambda_1&0&0&0\\0&\lambda_2&0&0\\0&0&\lambda_3&0\\0&0&0&\lambda_4
	\end{array}\right),\label{15}\\
	R&=\left(\begin{array}{cccc}0&2g&g^2-\omega \delta-T(H)&g^2-\omega \delta+T(H)\\0&2g&-g^2-\omega \delta+T(H)&-g^2-\omega \delta-T(H)\\0&-\delta&-2gH&-2gH\\1&0&-g\delta&-g\delta
	\end{array}\right).
	\end{align}

We multiply $R$ with $M$ to construct a more general diagonal matrix:
	\begin{align}\label{17}
	S=RM,~~
	M=\left(
	\begin{array}{cccc}
	1 & 0 & 0 & 0 \\
	0 & 1 & 0 & 0 \\
	0 & 0 & \beta  & 0 \\
	0 & 0 & 0 & \alpha (H) \\
	\end{array}
	\right),
	\end{align}	
where	$\beta$ is a constant number, and $\alpha(H)$ is a function about $H$ and $\beta$, which we will solve later.	Then we have		
\begin{eqnarray}\label{19}
S=\left(
\begin{array}{cccc}
0 & 2 g & \beta  \xi(H) &\eta(H)\alpha (H) \\
0 & 2 g & \beta   \tau(H) & \kappa(H) \alpha (H) \\
0 & -\delta  & \beta  \gamma (H) & \gamma (H)\alpha (H)  \\
1 & 0 & -g \delta\beta    & -g \delta  \alpha (H) \\
\end{array}
\right)
\end{eqnarray}	
where $\gamma (H)=-2 g H,  \xi(H)=g^2-\omega \delta -T(H),   \tau(H)=-g^2-\omega \delta +T(H), \kappa(H)=-g^2-\omega \delta -T(H),    \eta(H)=g^2-\omega \delta +T(H)$. We can find that
		\begin{eqnarray}\label{21}
		H(\sigma^+ a,\sigma^- a^\dagger,\sigma_z,1)S=(\sigma^+ a,\sigma^- a^\dagger,\sigma_z,1)S\Lambda.
		\end{eqnarray}
Then, from the technique in \cite{ge2000} we obtain
\begin{eqnarray}\label{22}
(1,D,b,b^+)=(\sigma^+ a,\sigma^- a^\dagger,\sigma_z,1)S,
\end{eqnarray}
such that		
		\begin{align}\label{23}
		b^+&=\sigma^+ a\left[g^2+T(H)-\omega \delta \right]\alpha (H)+\sigma^- a^\dagger\nonumber\\
            &~~~~~~~~~~\times\left[-g^2-T(H)-\omega \delta \right]\alpha (H)\nonumber\\
            &~~~~~~~~~~+\sigma_z\gamma (H)\alpha (H) -\text{g$\delta \alpha $}(H),\\
		b&=\sigma^+ a\left[g^2-T(H)-\omega \delta \right]\beta+\sigma^- a^\dagger\nonumber \\
&~~~~~~~~~~\times\left[-g^2+T(H)-\omega \delta \right]\beta\nonumber\\
&~~~~~~~~~~+\sigma_z \beta  \gamma (H)-\text{g$\delta \beta $}.
		\end{align}
Here $D$ is equivalent to $C$. From Eqs. (\ref{21}) and (\ref{22}), we have $H(1,S1,b,b^+)=(1,S1,b,b^+)\Lambda$, and
	\begin{equation}\label{MorseHb}
	[H,b]=b(\lambda_3-H), \;\; [H,b^+]=b^+ (\lambda_4-H),
	\end{equation}
which recover the definitions of raising and lowering operators.
So the operators $b^+$ and $b$ are the raising and lowering operators of the Hamiltonian operator $H$, respectively.
		
In what follows we construct $\alpha(H)$ to make the raising and lowering operators mutually adjoint
	\begin{equation}\label{mj}
	(b^+)^\dag=b.
	\end{equation}	
Suppose $F(H)$ is a real function of $H$, we have from Eq. (\ref{matrix-tech}) the following operator equation
\begin{eqnarray}\label{27}
F(H)(\sigma^+ a,\sigma^- a^\dagger,\sigma_z,1)=(\sigma^+ a,\sigma^- a^\dagger,\sigma_z,1)R F(\Lambda) R^{-1},
\end{eqnarray}
and, from Eq. (\ref{mj})	and  Eq. (\ref{27}),
\begin{eqnarray}\label{28}
\alpha(H)=\beta(1+\frac{2 g^2}{T(H)}).
\end{eqnarray}	
This is the relation between $\alpha(H)$ and $\beta$ .

\section{Determination of energy spectrum and wave functions}		

We will use the raising and lowering operators to determine
its energy spectrum and wavefunctions.	

\begin{enumerate}
  \item Ground state: For the ground state $|\psi_0\rangle$ and the zero-point energy  $E_0$ , it must be satisfied that
	\begin{eqnarray}\label{29}
	b|\psi_0\rangle=0,~~~
	H|\psi_0\rangle=E_0|\psi_0\rangle.
	\end{eqnarray}
Suppose in the computation basis that
\begin{eqnarray}\label{31}
|\psi_0\rangle=	\left(
\begin{array}{c}
\sum\limits_{n=0}^\infty e_n|n\rangle\\
\sum\limits_{n=0}^\infty f_n|n\rangle \\
\end{array}
\right).
\end{eqnarray}
From Eqs.(29) and (31),
for convenience, we introduce the notations
\begin{align}\label{32}
\alpha_1(E_0)&=\gamma (E_0)-g\delta,\nonumber\\
\alpha_2(E_0)&=g^2-\omega \delta -T(E_0),\nonumber\\
\beta_1(E_0)&=-g^2-\omega \delta +T(E_0),\nonumber\\
\beta_2(E_0)&=-\gamma (E_0)-g\delta,
\end{align}
and so
\begin{eqnarray}
\alpha_1\sum\limits_{n=0}^\infty e_n|n\rangle+\beta_1\sum\limits_{n=0}^\infty \sqrt{n+1}f_{n+1}|n\rangle=0,\label{33}\\
\alpha_2\sum\limits_{n=0}^\infty \sqrt{n+1}e_n|n+1\rangle+\beta_2\sum\limits_{n=0}^\infty f_n|n\rangle=0.\label{34}
\end{eqnarray}	
Through analysis of Eqs.(\ref{33}) and (\ref{34}), we find that if
$e_n=0,\beta_2=0,\beta_1=2g^2$,$f_n=\left\{\begin{array}{l}
1,n=0	\\
0,n\neq0
\end{array}\right..$
We obtain the ground state
\begin{equation}
\label{44}
\left\{\begin{array}{l}
|\psi_0\rangle=	\left(
\begin{array}{c}
0\\
|0\rangle \\
\end{array}
\right),	\\
E_0=\frac{\delta}{2}+\omega.
\end{array}\right.
\end{equation}	
%This is a special isolated single state,in addition to ,other states  belong to the state of the two-level system.in other words, there are two eigenstates in the two-dimensional subspace can be labeled by $+$ and $-$.	

  \item Excitation states:
  We have from Eqs.(\ref{33}) and (\ref{34}) that
\begin{eqnarray}\label{46}
\begin{array}{llll}
b|\psi_n^-\rangle=0,~~~
H|\psi_n^-\rangle=E_n^-|\psi_n^-\rangle,
\end{array}
\end{eqnarray}	
where $n = 1,2,3,4 \cdots$, and so	
\begin{eqnarray}
\label{47}
\left\{\begin{array}{l}
|\psi_n^-\rangle=\sin\theta_n	|g,n+1\rangle-\cos\theta_n|e,n\rangle\\\\
E_n^-=\omega(n+1)-\sqrt{g^2(n+1)+\frac{\delta^2}{4}}
\end{array}\right..
\end{eqnarray}	
	where $\tan\theta_n=\frac{-\delta+\sqrt{\delta^2+4g^2(n+1)}}{2g\sqrt{n+1}},n = 1,2,3,4 \cdots$	
	
Similarly, from
\begin{eqnarray}\label{48}
\begin{array}{lllll}
b^+|\psi_n^-\rangle=\chi|\psi _n^ +\rangle,~~~
H|\psi _n^ +\rangle=E_+|\psi _n^ +\rangle,
\end{array}
\end{eqnarray}
we have
\begin{align}
\label{49}
%\left\{
|\psi _n^ +\rangle=\cos\theta_n	|g,n+1\rangle+\sin\theta_n|e,n\rangle,\nonumber\\
E _n^ +=\omega(n+1)+\sqrt{g^2(n+1)+\frac{\delta^2}{4}}.
%\right.
\end{align}

\end{enumerate}

To summarize, the energy spectra and the wavefunctions are
\begin{align}
\label{50}
\begin{array}{l}
|\psi_0\rangle=	\left(
\begin{array}{c}
0\\
|0\rangle \\
\end{array}
\right),~~~
E_0=\frac{\delta}{2}+\omega(n+1),
\end{array}
\end{align}
for the ground state, and
\begin{align}
|\psi_n^+\rangle=\cos\theta_n	|g,n+1\rangle+\sin\theta_n|e,n\rangle,\nonumber\\
|\psi_n^-\rangle=\sin\theta_n	|g,n+1\rangle-\cos\theta_n|e,n\rangle,\nonumber\\
E_n^\pm=\omega(n+1)\pm\sqrt{g^2(n+1)+\frac{\delta^2}{4}},
\end{align}	
for the excitation states.

\section{Algebraic structure of the JCM}

We redefine three generators $J_0$ and $J_\pm$ with $H$, $b^+$ and $b$ satisfying
 commutation relations
	\begin{eqnarray}\label{51}
	[H,b]=-b f(H),   \ \ \  [H,b^+]=f(H)b^+, \label{commut}
	\end{eqnarray}
where $	f(H)=-\frac{g^2-T(H)}{\omega}.$
 From Eq.(\ref{commut}) we also have
 \begin{eqnarray}
 \begin{array}{lllll}
 Hb=b(H-f(H))=b\lambda_3=b\frac{g^2+H\omega-T(H)}{\omega},
 \end{array}	
 \end{eqnarray}
or more generally,
\begin{equation}
\begin{split}
 F(H)b&=bF(\lambda_3)=bF(H-f(H))\\
&=bF(\frac{g^2+H\omega-T(H)}{\omega}).
\end{split}
\end{equation}
Thus,
\begin{eqnarray}
[\frac{1}{2g^2}T(H),b]=-b,~~~
[\frac{1}{2g^2}T(H),b^+]=b^+.
\end{eqnarray}

Then we have
\begin{eqnarray}\label{54}
\left[ J_0 , b \right]=-b,~~
\left[ J_0, b^{\dag} \right] = b^{\dag},~~J_0=\frac{1}{2g^2}T(H)+\nu.
\end{eqnarray}
Furthermore,
\begin{widetext}
\begin{align}\label{55}
J_{-}&=b\xi(J_0), ~~~ J_{+}=\xi(J_0)b^{\dag},~~~
\left[ J_{+} , J_{-} \right]  = b^+ b \;
\xi^2(J_0)-bb^+\xi^2(J_0+1),
\end{align}
and
\begin{align}
	b^\dagger b&=f(H)\nonumber\\
&=\left( {H - \frac{\omega }{2} - \omega C} \right)\frac{{2T(H)\left( {{g^2} + 2H\omega  - T(H)} \right)}}{\omega } + \frac{1}{\omega }\left( {2{g^4}H + {\delta ^2}\omega \left( {2H\omega  - T(H)} \right) - 2{g^2}H\left( { - 4H\omega  + T(H)} \right)} \right)\nonumber\\
	&=f(J_0)\nonumber\\
&=\frac{{\left( {{J_0} - \nu } \right)\left( {{g^4}{{\left( {1 - 2{J_0} + 2\nu } \right)}^2} - {\delta ^2}{\omega ^2}} \right)\left( {{g^4}{{\left( {1 + 2{J_0} - 2\nu } \right)}^2} - \left( {\left( {2 + 4C} \right){g^2} + {\delta ^2}} \right){\omega ^2}} \right)}}{{2{g^2}{\omega ^2}}},\\
	bb^\dagger&=g(H)\nonumber\\
&=- \left( {H - \frac{\omega }{2} - \omega C} \right)\frac{{2{{\left( {2{g^2} + T(H)} \right)}^2}\left( {{g^2} + 2H\omega  + T(H)} \right)}}{{\omega T(H)}}\nonumber\\
	&~~~~~~+  \frac{{\left( {2{g^2} + T(H)} \right)\left[ {6{g^4}H + {\delta ^2}\omega \left( {2H\omega  + T(H)} \right) + 2{g^2}\left( {4{H^2}\omega  + {\delta ^2}\omega  + 3HT(H)} \right)} \right]}}{{\omega T(H)}}\nonumber\\
&=g( J_0 )\nonumber\\
	&= - \frac{{{{\left( {1 + {J_0} - \nu } \right)}^2}\left( {{g^4}{{\left( {1 + 2{J_0} - 2\nu } \right)}^2} - {\delta ^2}{\omega ^2}} \right)\left( {{g^4}{{\left( { - 1 + 2{J_0} - 2\nu } \right)}^2} - \left( {\left( {2 + 4C} \right){g^2} + {\delta ^2}} \right){\omega ^2}} \right)}}{{2{g^2}\left( {{J_0} - \nu } \right){\omega ^2}}},\\
{\xi ^2}\left( {{J_0}} \right) &= \frac{{2{g^2}{\omega ^2}}}{{\left( {{J_0} - \nu } \right)\left( {{g^4}{{\left( { - 1 + 2{J_0} - 2\nu } \right)}^2} - {\delta ^2}{\omega ^2}} \right)}}\nonumber.
	\end{align}

So
	\begin{align}\label{44}
\left[ {{J_ + },{J_ - }} \right] = {g^4}{\left( {1 + 2{J_0} - 2\nu } \right)^2} - \left( {\left( {2 + 4C} \right){g^2} + {\delta ^2}} \right){\omega ^2} + \frac{{\left( {1 + {J_0} - \nu } \right)\left( {{g^4}{{\left( {1 - 2{J_0} + 2\nu } \right)}^2} - \left( {\left( {2 + 4C} \right){g^2} + {\delta ^2}} \right){\omega ^2}} \right)}}{{\left( {{J_0} - \nu } \right)}}.
	\end{align}

\end{widetext}

\section{Bell nonlocality}
We introduce ``pseudo-spin'' operators~\cite{ChenZB2002}
%
%\begin{eqnarray}
%\left\{ \begin{array}{l}
%{s_z} = \sum\limits_{n = 0}^\infty  {\left( {\left| {2n + 1} \right\rangle \left\langle {2n + 1} \right| - \left| {2n} \right\rangle \left\langle {2n} \right|} \right)} \\
%{s_ - } = \sum\limits_{n = 0}^\infty  {\left| {2n} \right\rangle \left\langle {2n + 1} \right| = {{\left( {{s_ + }} \right)}^\dag }}
%\end{array} \right.
%\end{eqnarray}
%And corresponding to the  "pseudospin" Pauli operators:
\begin{eqnarray}
\left\{ \begin{array}{l}
{{\tilde \sigma }_x} = \sum\limits_{n = 0}^\infty  {\left( {\left| n \right\rangle \left\langle {n + 1} \right| + \left| {n + 1} \right\rangle \left\langle n \right|} \right)}, \\
{{\tilde \sigma }_y} = i\sum\limits_{n = 0}^\infty  {\left( {\left| {n + 1} \right\rangle \left\langle n \right| - \left| n \right\rangle \left\langle {n + 1} \right|} \right)}, \\
{{\tilde \sigma }_z} = \sum\limits_{n = 0}^\infty  {\left( {\left| n \right\rangle \left\langle n \right| - \left| {n + 1} \right\rangle \left\langle {n + 1} \right|} \right)},
\end{array} \right.
\end{eqnarray}
%When $n=0$,the  "pseudospin" Pauli operators become  Pauli operators.
and we define the Bell operator~\cite{CHSH69} as
\begin{eqnarray}
\begin{array}{lllll}
{B_{CHSH}}&=& \left( {\hat {\vec \sigma}  \cdot \vec a} \right) \otimes \left( {\tilde {\vec \sigma}  \cdot \vec b} \right) + \left( {\hat {\vec \sigma}  \cdot \vec a} \right) \otimes \left( {\tilde {\vec \sigma}  \cdot \vec b'} \right) \\
&+& \left( {\hat {\vec \sigma}  \cdot \vec a'} \right) \otimes \left( {\tilde {\vec \sigma}  \cdot \vec b} \right) - \left( {\hat{ \vec \sigma}  \cdot \vec a'} \right) \otimes \left( {\tilde {\vec \sigma}  \cdot \vec b'} \right),
\end{array}
\end{eqnarray}
where $\vec a,\vec a',\vec b,\vec b'$ are four unit vectors, $\hat {\vec \sigma}=(\hat\sigma_x,\hat\sigma_y,\hat\sigma_z)$ denotes the usual Pauli matrix (i.e., by taking $n=0$ in the pseudo-spin operators).

 We can construct a test of Bell's inequality with the excitation states, for example, the $| {\psi _n^ - }\rangle$ for a certain $n$:
\begin{eqnarray}
\left| {\psi _n^ - } \right\rangle & =& \sin {\theta _n}\left| {g,n + 1} \right\rangle  - \cos {\theta _n}\left| {e,n} \right\rangle\nonumber \\
&\equiv& \sin {\theta _n}\left| 1 \right\rangle \left| {n + 1} \right\rangle  - \cos {\theta _n}\left| 0 \right\rangle \left| n \right\rangle
\end{eqnarray}
with $|0\rangle=(1,0)^{\rm T},~|1\rangle=(0,1)^{\rm T}$. By taking
\begin{equation}
  \begin{split}
   & \vec a=(0,0,-1),~~~ \vec a'=(1,0,0),\\
   & \vec b=(-\sin\theta^*,0,\cos\theta^*),\\
   & \vec b'=(\sin\theta^*,0,\cos\theta^*),\\
   & \theta^*=\frac{\pi}{2}+\arctan\frac{1}{\sin2\theta_n},
  \end{split}
\end{equation}
it turns out that $\langle B_{CHSH}\rangle=2\sqrt{1+\sin^2(2\theta_n)}$, which violates the local hidden variable bound 2, except for $\theta_n=0,\pi/2$. This shows the ubiquitous presence of Bell nonlocality in the JCM.

%\[\left| 0 \right\rangle  = \left( {\begin{array}{*{20}{c}}
%	1\\
%	0
%	\end{array}} \right),\left| 1 \right\rangle  = \left( {\begin{array}{*{20}{c}}
%	0\\
%	1
%	\end{array}} \right)\]

\section{Discussion}	

In this paper, we have obtained the raising and lowering operators for the JCM by means of the matrix-diagonalizing technique, and then worked out the energy spectra and wavefunctions in the computation basis. We have then revealed the dynamical symmetry of the model by writing these shift operators in terms of generators of the Lie algebra. Finally, we have shown that Bell nonlocality is found to exist in the excitation states of the JCM, further justifying the merits of solving the JCM with the method used in the present paper.

\section{acknowledgments}

J.L.C. is supported by National Natural Science Foundations of China (Grant No. 11475089). F.L.Z. is supported by National Natural Science Foundations of China (Grant No. 11675119 and No. 11575125)

\end{document}